\documentclass{bmcart}

\usepackage{amsthm,amsmath}
\usepackage[utf8]{inputenc} 
\usepackage{hyperref}
\usepackage{graphicx}

\startlocaldefs
\endlocaldefs

\begin{document}

\begin{frontmatter}

\begin{fmbox}
\dochead{Letter}


\title{Notes on Cloud computing principles}


\author[
   addressref={aff1,aff2},                   
   corref={aff1},                       
   email={thomas.e.sandholm@hp.com}   
]{\inits{TS}\fnm{Thomas} \snm{Sandholm}}
\author[
   addressref={aff2},
   email={dlee@kaist.ac.kr}
]{\inits{DL}\fnm{Dongman} \snm{Lee}}


\address[id=aff1]{
  \orgname{HP Labs}, 
  \street{1501 Page Mill Rd},                     %
  \postcode{94304},                                
  \city{Palo Alto, CA},                              
  \cny{USA}                                    
}
\address[id=aff2]{%
  \orgname{Department of Computer Science, KAIST},
  \street{291 Daehak-ro},
  \postcode{305-701}
  \city{Daejeon},
  \cny{Korea}
}



\end{fmbox}


\begin{abstractbox}

\begin{abstract} 
This letter provides a review of  
fundamental distributed systems and economic 
Cloud computing principles. These principles
are frequently deployed in 
their respective fields, but their 
interdependencies are often neglected.
Given that Cloud Computing first and foremost is 
a new business model, a new model to sell
computational resources, the understanding of these
concepts is facilitated by treating them in unison. Here, we review
some of the most important concepts and how they
relate to each other.
\end{abstract}




\end{abstractbox}
%

\end{frontmatter}



\section*{Introduction}
Imagine that you have to go on a trip to meet a friend in a different city.  There are many modes of 
transportation available to you. You can drive there by car, take a taxi, share a ride in a van, take a 
bus or a train, or even fly there in an airplane.  Your choice is determined by your general preference 
for these options. In particular, your choice depends on the economics and convenience of these alternatives 
given the characteristics of the trip, including distance to destination and time available. The cost of 
the choice you make in turn is related to how many other people are sharing the same mode of transportation, 
and how expensive it is to operate the transportation vehicle and infrastructure.

Now compare this choice to the choice of energy supplier that people faced in the early 20th century.  
You could buy your own electric generator, but it was not very cost efficient if your needs varied diurnally or 
seasonally. As it became apparent that electricity was as invaluable of a commodity as gas, water and the telephone, 
utility companies and national electrical grids that could aggregate and distribute electricity on demand replaced
the privately owned generators.

Cloud computing~\cite{armbrust2010} could be seen as an effort to commoditize computing and distribute and operate 
it as efficiently as the electrical grid while still offering consumers the plethora of alternatives known from 
the transportation domain. The pre-cloud era could be compared to everyone driving around in their own car and 
using their own generators. The cloud era allows computing to be used similarly to public transportation and 
makes it possible to tap into computing power with the same ease that you plug in your appliances to the electrical 
grid at home.
To distinguish the Cloud from its predecessors it is often defined as a use of computing resources that are 
delivered as a service over a network.  The way in which you provision these services holds the key to the innovation.

Cloud services need to be scalable, fault-tolerant, highly available, high-performance, reliable and easy to use, 
manage, monitor and provision efficiently and economically.  One early realization by Cloud computing pioneers was
that meeting all these requirements for services handling massive amounts of data and huge numbers of concurrent users 
called for innovation in the software stack as opposed to the highly specialized hardware layer. The hardware is reduced to a 
commodity and the Quality of Services (QoS) are instead provided by a fully integrated and hardware agnostic software stack. 
Virtualization became the new silver bullet.

As the demand for compute power increased with more users coming on-line and more data being published on-line it became apparent 
that some drastic architectural changes had to be introduced to provision compute resources more efficiently.  The most 
prominent enabler for efficient resource provisioning was data center consolidation.  Instead of using spare cycles 
from arbitrary privately owned nodes in a network~\footnote{as was done in many P2P networks at the time}, it was more cost 
effective to provide high QoS by consolidating computing in highly streamlined data centers packed with low-cost 
dedicated compute and storage clusters in a highly reliable and fast network.  These data centers were also frequently 
deployed in areas where energy and labor were cheap to further cut operational costs.

Data-center consolidation and more aggressive sharing of compute resources lead to the following key benefits of Cloud computing:
\begin{enumerate}
\item{Lower cost of using compute resources}
\item{Lower cost of provisioning compute resources}
\item{Reduced time-to-market}
\end{enumerate}

The first benefit can be attributed to only paying for the resources when you use them. When you do not use them the provider 
can allocate them to other users. Being able to host multiple users or tenants on the same infrastructure allows the provider 
to utilize the resources more efficiently and thereby increase the return on investment (ROI).  This win-win relationship between 
compute users and providers is the reason why most companies switch to Cloud architectures. The growth and sudden popularity of Cloud 
computing was, however, not fueled by traditional, established companies.  Start-ups were the pioneering users of Cloud 
technology as it reduced their time-to-market and provided them with less up-front risk to stand up a demo or beta version. 
If the users did not flock, not much harm was done, you just stopped paying for the resources. If there was an unexpected flash crowd of 
people bombarding the service, you would just pay for more resources. This type of usage is often referred to as the elasticity of the Cloud. 
The Cloud allows you to scale down as easily and as quickly as you scale up.

Below we will review some of the fundamental concepts of distributed computing at scale, and then
relate these concepts to economic principles that help us understand the trade-offs
governing their deployment. The main motivation for studying these
economic principles is that solely maximizing systems metrics, such as, throughput, 
response time and utilization may not always be the most profitable strategy for a Cloud provider.

Before delving into these principles we will first take a look back at technologies that predated
Cloud computing to see how the architecture of this new computing paradigm evolved into 
its current state.

\section*{Historical evolution}
The vision of organizing compute resources as a utility grid materialized in the 1990s as an effort to solve 
grand challenges in scientific computing. The technology that was developed is referred to as Grid Computing~\cite{kesselman1998}, 
and in practice involved interconnecting high-performance computing facilities across universities in regional, national, and pan-continent Grids.  
Grid middleware was concerned with transferring huge amounts of data, executing computational tasks across administrative domains, and 
allocating resources shared across projects fairly. Given that you did not pay for the resources you used, but were granted them based on your 
project membership, a lot of effort was spent on sophisticated security policy configuration and validation.  The complex policy landscape that ensued
hindered the uptake of Grid computing technology commercially. Compare this model to the pay-per-use model of Cloud computing and it then becomes 
easy to see what, in particular, smaller businesses preferred. Another important mantra of the Grid was that local system administrators should 
have the last say and full control of the allocation of their resources. No remote users should have full control or root access to the expensive 
super computer machines, but could declare what kind of software they required to run their jobs.  Inherently in this architecture is the notion of 
batch jobs. Interactive usage or continuous usage where you installed, configured and ran your own software, such as a Web server was not 
possible on the Grid. Virtual machine technology~\cite{barham2003} released the Cloud users from this constraint, but the fact that it was 
very clear who pays for the usage of a machine in the Cloud also played a big role. In summary, these restrictions stopped many of the Grid protocols 
from spreading beyond the scientific computing domain, and also eventually resulted in many scientific computing projects migrating to Cloud technology.

Utility computing~\cite{rolia2002} refers to efforts in the industry around the turn of the millennium to improve manageability and 
on-demand provisioning of compute clusters. At this time, companies were very skeptical to running their confidential workloads off 
premise and thus utility computing was often sold on a cluster-by-cluster basis and installed on a company-by-company or organization-by-organization basis. 
This deployment model made it very expensive to get up and running, which ironically had been one of the key claimed benefits of utility computing.  
Nevertheless, it started to become clear around this time that virtualization was the key to on-demand provisioning of compute resources.
Web services and Service-Oriented Architectures~\cite{gottschalk2002} were touted as the solution to many of the problems seen in the earlier 
efforts of Utility and Grid computing. Providing a standard API would allow infrastructure to be allocated programmatically based on demand.  
The APIs and protocols were borne out of the evolution of the World Wide Web (WWW)  that started to provide more dynamic and interactive
content on Web pages leading to the phenomenon of mashups.  Mashups in the early days essentially scraped HTML from various Web pages to 
dynamically create a value-adding service on a new Web page. As this was error prone it was quickly realized that APIs were needed and 
the first Web services protocols, such as SOAP~\cite{box2000}, were designed.

By the time Amazon launched their Elastic Compute Cloud (EC2) service in 2006, both Web service APIs and virtualization technology (e.g. Xen[3]) 
had matured enough to form a compelling combination or a perfect storm to deliver the first real public utility computing service that had been 
envisioned a decade earlier.

In summary, the vision of the Grid combined with Virtual Machine technology and Web service APIs were the essential characteristics of the first Clouds.
Next, we will review the fundamental distributed systems principles underlying today's Cloud systems. 

\section*{Computational principles}
\subsection*{Multi-tenancy}
A tenant in the Cloud context is a user of Cloud infrastructure, i.e. Infrastructure-as-a-Service (IaaS) services~\cite{lenk2009}.  
A VM owner is an example of a tenant and if multiple VM owners are allocated on the same physical machine it is an example of multi-tenancy~\cite{wilder2012}. 
The difference between a multi-(end)-user service and a multi-tenant service is that a multi-user offering may benefit 
from having users know about each other and explicitly share social content to promote the network effect. 
A multi-tenant solution could internally benefit from shared physical resources but must give the impression of an 
exclusive offering to each of the tenants. As an example, hosting the Facebook service on a Web server in the Cloud would be 
an example of a multi-user service, but hosting both a Twitter Web server and a Facebook Web server in the same Cloud data center would 
be an example of multi-tenancy. From this definition it is clear that the IaaS provider needs to provide mechanisms to isolate the tenants from each other.

Multiple tenants need to be isolated in terms of privacy, performance and failure:

\begin{itemize}
\item{\bf Privacy Isolation.} Multiple tenants must not have access to each other's data. This may seem like an easy requirement 
to meet but in a typical file system there may be traces left after a file even after removing it, which would violate this property.
\item{\bf Performance Isolation.} Multiple tenants must not be effected by each other's load. If one tenant starts running a CPU 
intensive task and other tenants see a drop in performance as a result, then this property is violated.
\item{\bf Failure Isolation.} If a tenant either inadvertently or maliciously manages to crash its compute environment it 
should not effect the compute environment of other users. Imagine a Java VM hosting multiple applications such as a Tomcat Servlet engine. 
Now, if one servlet Web app crashes the VM, then the other apps in the same VM would also crash. This failure would in that case be a violation of the failure isolation property.
Virtual machines offer a popular technique to ensure isolation, but in some cases the overhead of virtualization, of e.g. IO and network, is 
too high so a trade-off has to be made between isolation level and performance.
\end{itemize}
Ensuring these levels of isolation is closely related to the strategy used to allocate resources to tenants, which we will discuss next.

\subsection*{Statistical multiplexing}
One major benefit related to data center consolidation that we discussed in the introduction is statistical-multiplexing~\cite{knightly1999}. 
The idea behind statistical multiplexing is that bursty workloads that are consolidated on the same Cloud infrastructure may in 
aggregate display a less bursty pattern.  Figure~\ref{fig:statmult} shows an example of statistical multiplexing with two workloads 
exhibiting complementing demand over time. 
\begin{enumerate}
\item	Without an elastic Cloud infrastructure, the most common way of provisioning resources to tenants is to allocate resources that meet the peak demand of each workload. Clearly, this leads to a major waste in resources for the majority of the time. Statistical multiplexing allows an allocation that is substantially lower than the sum of the peaks of the workloads.
\item 	Ideally if statistical multiplexing is applied on a large number of independent workloads, the aggregate will be stable, i.e. a straight line in the demand chart. If this is the case, it is enough to just allocate the sum of the averages of resource demand across all workloads.
\item 	Now assuming that we are in an elastic Cloud environment and we can slice resource allocations by time akin to how an OS time-shares CPU between processes. In this scenario further reductions in resource allocations may be achieved by simply allocating the sum of resource demand across all workloads in each time slice.
\item	Finally if each time slice only has a single workload active at any point in time, the allocation reduces to just the maximum demand across the workloads. 
\end{enumerate}
 
This model of perfect statistical multiplexing is hard to achieve in practice. The main reason for this is that workloads tend to be correlated. 
The effect is known as self-similarity.  Self-similar workloads have the property that aggregating bursty instances will produce an 
equally bursty aggregate, something that is often observed in practice.  However, there are many techniques to recreate the effects of 
statistical multiplexing without having to hope for it to occur organically. For instance you could measure the correlation between 
workloads and then schedule workloads that are complementing on the same resources. These techniques are sometimes referred to as 
optimal packing of workloads or interference minimization~\cite{delimitrou2014}. Poor statistical multiplexing tends to lead to
low utilization, or unmet demand, as we will discuss further when we review the economic principles governing under and over-provisioning. 

\subsection*{Horizontal scalability}
An application or algorithm that runs in the Cloud will not be able to scale up and down with the 
infrastructure unless it can run at least in part in parallel. Execution in the Cloud requires efficient 
scaling across machines, referred to as horizontal scalability. A local program running on a single machine 
on the other hand only needs to scale vertically, i.e. run faster as local resources such as CPU, memory, and disk are added.  
How well a program scales is thus related to the parallelizability of its algorithms. This effect is formalized in 
what is called Amdahl's Law~\cite{amdahl1967}:
\begin{equation}
T(n) = T(1)(B+(1-B)/n)
\end{equation}
Amdahl's Law predicts the expected speed-up of a program or algorithm when run over multiple machines. 
$T(n)$ is the time taken to run on $n$ machines. $B$ is the fraction of the program that needs to run serially, 
i.e. that cannot be parallelized. Note that several disjoint sections in the execution path may need to run serially 
to collect, distribute or synchronize parallel computations.  It is clear that minimizing $B$ maximizes the speedup.  
However, the most important consequence of Amdahl's Law is that it sets a theoretical cap on how many machines a program 
will benefit from running on, beyond which point adding new machines will not make the program run faster. If $B$ is close to 
negligible we can expect {\it linear scalability}.  Adding $x$ machines will make the program run $x$ times faster.  If the program 
speedup grows at a slower rate than the number of machines added, which is the common case due to various overheads of distribution, we 
refer to {\it sublinear scalability}. The program may also speedup at a faster rate than the machines being added, 
in which case the program is said to exhibit {\it superlinear scalability} (see Figure~\ref{fig:scalability}). This effect may happen if there is some common resource 
like a shared cache that benefits from more usage, e.g., more cache entries and fewer cash misses.
 
\subsection*{Data partitioning}
To achieve data scalability, i.e. scalable access to data, it is common to not only replicate and distribute individual data 
items but also to replicate the database instances.  This design is very popular in Cloud data centers and is known as a 
sharding or shared-nothing architecture~\cite{stonebraker1986}.  Each shard or distributed database instance is responsible 
for a subset of the data items or rows in a traditional RDBMS. The notion of shared-nothing comes from the fact that each shard 
is a self-sufficient denormalized store capable of handling all requests to the database for its allocated subset of items independently. 
By banning merges between shards as is typical in RDB joins one can ensure efficient data-parallel delivery of typically very large items. 
Now, the main question is how to determine which item is stored in which shard. This problem is known as data partitioning and there are 
two general solution strategies. The data can either be partitioned by range or by hashing.  In range partitioning, the key range, e.g.  
an unsigned integer from 0 to $2^{32}-1$,  is split into equally sized intervals and each interval is mapped to a shard. That shard is then responsible 
for delivering the values for all keys in that range.  The advantage of this type of partitioning is that keys that are often accessed together 
may be collocated on disk or cached in the same memory block and thereby be retrieved very efficiently during range queries.  For instance, if 
the range denotes geographic locations, it may be used as a technique to query all information in a geographic area, or if the range denotes time it could be 
used to get a history of all items in a time window. The downside of this technique is that it assumes that the keys are uniformly balanced across the 
key range, which is rarely the case.  If fast access to individual items and perfectly balanced partitions to accomplish a higher level of data 
parallelism is a priority, then hash-based partitioning is more appropriate. Instead of mapping the shards directly to key ranges they are mapped to 
ranges of hashes of the keys. As long as the shards stay alive and can serve requests for their hash partition reliably, this technique results in 
good load-balancing in practice.  Next we will discuss a refinement of basic hash partitioning when the shards are unreliable, i.e. may go down and come back up dynamically,
e.g. based on elastic Cloud allocations to scale-up or down to meet demand. 
In summary, range partitioning and hash partitioning are efficient for partial key scans and individual key lookups respectively (see Figure~\ref{fig:range} and \ref{fig:hash}).

\subsection*{Consistent hashing}
If we add a new node to meet more demand or if a node fails or is taken down due to low demand, then the 
hashed value,  $hash(x) \bmod n$, will change for almost all keys $x$ (see Figure~\ref{fig:consist}).  This change has the 
undesirable implication of having to move many of the values of x between the shards to keep the index up-to-date.  
Consistent hashing~\cite{karger1997} is a solution to this problem that tries to minimize the impact, i.e. keys that need to move 
between shards, if there is a change in shards serving the key values, i.e. the number $n$. The basic idea is to organize the hashed 
key space in a circle with values sorted in ascending order clockwise, known as the key ring, where the maximum and the minimum 
values are located adjacent to each other at the top.  Now, both keys and shard IDs are hashed into the same hash space in 
the key ring. To find the node that is responsible for storing the value of a key the key is first hashed and then located 
in the key ring.  The first shard ID encountered moving clockwise on the ring will be contacted to retrieve the value.  
This technique guarantees that only $|x| /n$ keys need to be re-mapped on average, where $|x|$ is the number of distinct keys, 
whereas traditional hashing required approximately $|x|$ re-mappings.  Today consistent hashing serves as the back-bone in 
many key-value stores, popular in Cloud data centers, as well as p2p overlays, a.k.a. distributed hash-tables (DHTs)~\cite{stoica2001}.
Hence, consistent hashing handles node elasticity efficiently in the Cloud.

To handle the case of node failure, and also to account for potential imbalance in the hashed key space, 
shards are typically added to the key ring multiple times. Shards may for instance duplicate key values stored in 
other shards located directly to the left in the key ring, as they would be responsible for serving the keys should these 
shards go down.  How many times a value is replicated, depends on the read and write semantics expected 
from the application, as well as the consistency guarantees given.  We discuss this trade-off between availability and consistency next. 

\subsection*{Eventual consistency}
Replication and duplication of data items in the Cloud is done to facilitate failover but also to achieve 
data-parallelism. In the latter case it is not unusual to have a very large number of replicas, in the extreme 
case one per node to allow each node to process the data concurrently.  Traditional consistency schemes known from 
the RDBMs field, such as 2PC and 3PC are known to scale poorly if the number of replicas is large and the nodes 
are unreliable. Hence, consistency in the Cloud is often relaxed, to not strictly follow the ACID properties of a 
traditional database. The window of inconsistency is the time between a successful update of a value until the 
time all reads of that value return the updated value. The size of this window that is allowed is application specific.  
Even within an application some updates that are more critical may need to have ACID (window size 0) guarantees whereas 
other values may allow a longer window of inconsistency. A series of techniques known as eventual consistency~\cite{vogels2009}, give a 
guarantee that an updated value will eventually show up in subsequent reads.  To understand the trade-offs the infrastructure has to make 
to provide these guarantees consider an application with a data item that has $N$ replicas that needs to write successfully to $W$ replicas for the whole 
update to succeed, and that needs to read from $R$ replicas to provide the requested consistency guarantee. In a system where $W+R> N$ it is clear that 
strict ACID-like guarantees with a window of inconsistency of 0 may be achieved.  In this case, the only trade-off is what the values of 
$W$ and $R$ should be. For a WORM (Write Once Read Many) workload $W$ would typically be greater than $R$, to allow for fast concurrent 
reads but potentially slow and unreliable writes. In a system with a high-frequency of updates, it may be worthwhile to reduce $W$ and as a 
consequence increase $R$. Note that any system can easily be turned into WORM by time-stamping each update and thereby making them immutable.
One example of a common configuration to achieve fault-tolerance is $N=3$,$W=2$,$R=2$ (see Figure~\ref{fig:eventual}).

Note that here it is allowed for one replica to fail to be updated as long as two nodes may be successfully read to maintain consistency.  
In general increasing $N$ and decreasing $W$ has the effect of increasing availability and thereby also data-parallelism, 
but reliability and consistency may be reduced. Increasing $R$ leads to less data parallelism but improves consistency guarantees. 
It is easy to see that allowing applications to specify $N$, $W$, and $R$ essentially allow them to configure the consistency and 
availability guarantees of the data items to be stored.  Typically in a practical application of eventual consistency the applications 
would hence just specify the availability and consistency guarantees desired, and the infrastructure would then map it to $N$, $W$ and $R$ 
values based on statistical guarantees (historically measured values).

Quorum consistency a.k.a. Paxos~\cite{lamport2001} consistency is a strict guarantee where $W$ and $R$ are both the majority set  of 
replicas. It is a technique that is used primarily for two practical reasons. First, it may be used to pick which is the new 
primary node in the case of replicas that are organized as a single primary and a pool of backups.  The second common use is as a 
distributed locking service.  During updates of a commonly used, replicated file you may want to ensure that all 
other access is blocked until the value is stored. According to the protocol, nodes that have voted in a 
quorum will need to hold off on voting on other values until the initial vote has been recorded. Voting here is 
analogous to approving an update and recording is analogous to updating.

In summary, eventual consistency makes a trade-off between performance, reliability and consistency.
Next we turn to the economic principles to understand the financial implications of various trade-offs made in system architectures. 

\section*{Economic principles}
\subsection*{Over and under provisioning}
As we alluded to in the section on statistical multiplexing, over-provisioning is a common strategy for allocating resources across tenants. 
Here we discuss the economic dilemma of over (Figure~\ref{fig:over}) versus under-provisioning (Figure~\ref{fig:under}) resources.
 
We can see that over-provisioning leads to a large area of idle resources over time. 
In financial terms this means high-operational cost, and lost opportunities to increase profit.  
To increase profit the IaaS provider may be tempted to lower the allocation to reduce the operational cost as 
seen in Figure~\ref{fig:under}. However, this leads to an even more severe drawback, unmet demand. 
Unmet demand means revenue loss, and can have long-term negative effects as customers who are denied access to a 
resource despite being willing to pay for it may not return. For this reason over-provisioning is more popular than 
under-provisioning. However, neither the IaaS provider nor the tenant may be able to perfectly predict the peaks, 
after all that is why they are running in the Cloud in the first place. In this case under-provisioning may occur inadvertently.

Hence, over-provisioning versus under-provisioning involves making a trade-off between profit and revenue loss.

\subsection*{Variable pricing}
Given all the issues of allocating resources to bursty demand, it is natural to ask whether this burstiness can be suppressed 
somehow as opposed to being accommodated. That is exactly the idea behind variable pricing or demand-driven pricing. 
The idea is to even out the peaks and valleys with incentives. If the demand is high we 
increase the price. This leads to tenants who cannot afford the higher price to back-off 
and thereby demand is reduced.  On the other hand, if the demand is low, a price drop may encourage tenants 
who would otherwise not have used some resources to increase their usage and thereby demand. The end result is a stable 
aggregate demand as in the statistical multiplexing scenario. The key benefits to IaaS providers include the ability to 
cash in on peak demand by charging premiums, and a mechanism to increase profit during idle times. Now, how can we ensure 
that the price is a good representation of demand? Here, microeconomic theory of supply and demand~\cite{mas1995} helps. 
 
If we plot the quantity of goods a supplier can afford to produce given a price for the good we get the supply curve. 
If we plot the quantity of goods requested by consumers given a price for the good we get the demand curve. 
The price at the point where the supply and demand curves meet is called the efficient marker price as it is a stable price that a 
market converges towards (see Figure~\ref{fig:supply}). To see why this is the case, consider the gray dot on the supply curve in Figure~\ref{fig:supply}. 
In this case the supplier observes a demand that is higher than the current quantity of goods produced. Hence, there is an opportunity for the 
supplier to increase the price of the good to afford to produce more goods to meet this demand.  Conversely, considering the black dot on the demand curve, we can see 
that the demand is higher than the volume of goods that the supplier can produce. In this case the demand will naturally go 
down and the consumers are likely to be willing to pay a higher price to get their goods. 

In general, variable pricing allows a provider to allocate resources more efficiently.

\subsection*{Price setting}
There are many ways to set prices for goods in a market. The most commonly known are various forms of 
auctions, spot prices and reservations.
In auctions, bidders put in offers to signal how much they are willing to pay for a good. 
In double actions, there are also sellers who put in asks denoting how much they are willing to sell the good for. 
The stock market is an example of a double auction. In computational markets, second price sealed bid auctions are 
popular since they are efficient in determining the price, i.e. reflect the demand, without too much communication. 
All bidders put in secret bids and the highest bidder gets the good for the price equalling the second highest bid.

In the case were there is not a completely open market price, that is there is just a single provider selling off compute resources, 
spot pricing is a common way of setting demand based prices. The spot price is computed on a running basis depending on the current level of demand. 
There could for instance be a base pay that is discounted or hiked based on demand fluctuations. A spot market differs from a 
futures market in that goods are bought and consumed immediately. Futures markets such as options are less common in practical 
computational markets today. 

Purchasing resources on a spot market involves a high risk of either having to pay more for the same allocation or being forced to 
reduce the allocation to stay within budget (see the section on predictability below). A common way to reduce the risk for users is 
to offer a reservation market. A reservation market computes the expected spot demand for some time in the future and adds a 
premium for uncertainty to arrive at a reservation price. Essentially you have to pay for the providers lost opportunity of 
selling the resources on the spot market. This way the risk is moved from to consumer of compute resources, i.e. the tenant, to the provider. 
If there is an unexpected hike in the demand and all resources have already been promised away in reservations there is 
no way for the provider to cash in on this demand, which constitutes a risk for the provider. 

In summary, reservation markets move the risk of uncertain prices from the tenant to the provider as uncertain demand.

\subsection*{The tragedy of the Commons}
The next principle we will discuss is a social dilemma referred to as the tragedy of the Commons~\cite{hardin1968}. 
The dilemma was introduced in a paper in 1968 by Garrett Hardin, where the following scenario was outlined. 

Imagine a public, government-owned piece of land with grass, in the UK referred to as a Common. 
Now, a number of shepherds own sheep that they need to feed on this Common to keep alive. 
The shepherds will benefit economically from the sheep because they can, for instance, sell their wool.  
Each shepherd faces the financial decision whether it would be more profitable to purchase another sheep 
to feed on the Common and extract wool for, or provide more food to each sheep by sticking with the current herd. 
Given that it is free to feed the sheep on the Common and the reduction in available food is marginal, it turns out that it is 
always optimal for a selfish shepherd trying to optimize his profit to buy another sheep. 
This has the effect of driving the Common into a slump where eventually no more grass is available and all sheep die 
and all shepherds go bankrupt.

One could argue that less selfish shepherds who are wary of the benefits of the group of shepherds as a prosperous community 
will not let the situation end in tragedy. However, there are many examples of communities that have gone extinct this way.  
In general what these communities have in common is that there is a high degree of free-riders, i.e. community members who take more 
from the common resources of the community than they give back.  Sometimes the effects are temporal and not as obvious since no one purposefully 
abuses the community. One example is the PlanetLab testbed~\cite{chun2003} used by systems researchers in the US.  
The testbed is distributed across a large number of organizations to allow wide area and large-scale experiments.  The weeks leading up to 
major systems conferences such as OSDI, NSDI, SOSP and SIGCOMM see extreme load across all machines in the testbed typically 
leading to all researchers failing to run their experiments. 

The opposite of free-riding is referred to as altruism. 
Altruists care about the community and are the backbone of a sustainable and healthy community. A good example of this is the Wikipedia 
community with a small (compared to readers) but very dedicated group of editors maintaining the order and quality of the information provided. 
The opposite of the tragedy of the Commons is the network effect where more users lead to greater benefits to the community, e.g. by 
providing more content as in the Wikipedia case.

The balance between free-riders and altruists as well as the regulations and pricing of resource usage determines whether 
the tragedy of Commons or the network effect prevails.

This concept is closely related to what economists refer to as externality~\cite{buchanan1962}, individual actions impose an
unforeseen positive or negative side-effect on the society. The archetypical example is
factory pollution. Such side-effects are mainly addressed in the Cloud by various infrastructure isolation
designs such as virtual machines, or virtual private networks (see discussion in the section on multi-tenancy
above).

\subsection*{Incentive compatibility}
One of the most frequently overlooked aspects of distributed systems is incentive compatibility~\cite{hurwicz1977}. Yet it is a property that all successful 
large-scale systems adhere to, the Cloud being no exception, and it is very often the main reason why proposed systems fail to take off.  
It is a concept borrowed from game-theory. In essence, an incentive compatible system is a system where it is in the interest of all 
rational users to tell the truth and to participate. In a systems context, not telling the truth typically means inserting 
incorrect or low quality content into the system to benefit your own interests. Incentive to participate is closely related to the notion of free-riding. 
If there is no incentive to contribute anything to a common pool of resources, the pool will eventually shrink or be overused to 
the point where the system as a whole becomes unusable. That is, the system has converged to a tragedy of the Commons. 
Ensuring that the system cannot be gamed is thus equivalent to ensuring that there is no free-riding and that all users contribute 
back to the community the same amount of valuable resources that they take out.  A new, untested, system with a small user base also has to 
struggle with a lack of trust, and in that case it is particularly important to come out favorable in the individual cost-benefit analysis, 
otherwise the potential users will just pick another system. Tit-For-Tat (TFT) is an example of an incentive 
compatible algorithm to ensure a healthy and sustainable resource sharing system.

If Cloud resources are sold at market prices it ensures incentive compatibility, .i.e. ensuring that the price is following the 
demand (in the case of a spot market) or the expected demand (in the case of a reservation market) closely has the effect of 
providing an incentive for both suppliers and consumers to participate in the market.  Earlier systems such as the Grid and P2P systems 
that did not have an economic mechanism to ensure incentive compatibility has historically had a much harder time of sustaining a 
high level of service over a long period of time due to frequent intentional and non-intentional free-riding abuses.
Hence, demand-based pricing helps ensure incentive-compatibility.

Computational markets that have demand-driven pricing may however still not be incentive compatible. If it for instance is very cheap to 
reserve a block of resources ahead of time and then cancel it before use, it could lead to an artificial spike in demand that could dissuade 
potential customers from using the resource.  This in turn would lead to the spot market price being lower, which could benefit the 
user who put in the original reservation maliciously.  In economic terms it is a classic example of someone not telling the truth (revealing their true demand in this case) 
in order to benefit (getting cheaper spot market prices). Another classic example is an auction where the bidders 
may overpay or underpay for the resource, just to make sure competitors are dissuaded to participate or to falsely signal personal demand. 

\subsection*{Efficiency}
Shared resource clusters such as the Grid, are commonly monitored and evaluated based on systems metrics such as utilization. 
A highly utilized system meant the resources typically funded by central organizations such as governments were being efficiently used. 
This type of efficiency is referred to as computational efficiency. It is a valuable metric to see whether there are opportunities 
to pack workloads better or to re-allocate resources to users who are able to stress the system more, i.e. a potential profit opportunity (see the section above on over and under provisioning). 
In a commercial system such as the Cloud it is also important to consider the value that the system brings to the users, because the 
more value the system brings to users the more they are willing to pay and the higher profit the Cloud provider is able to extract from a resource investment. 
This trade-off becomes apparent when considering a decision to allocate a resource to a user who is willing to pay \$0.1 an hour for some resource and 
utilize at close to 100\% versus another user who is willing to use the same resource over the same period of time but at 
90\% utilization and paying \$0.5 an hour. There is likely more idle time and unused resources if the second user is 
accommodated but the overall profit will be higher (0.5-0.1=\$0.4/hour).

To evaluate the economic efficiency~\cite{papadimitriou2001} one therefore often go beyond pure system metrics. 
In economics, utility functions are used to capture the preferences or the willingness of a user to pay for a resource.   
Maximizing the overall utility across competing users is then a common principle to ensure an overall healthy and sustainable ecosystem.  
This sum of utilities across all users is referred to as the {\it social welfare} of the system. To compare two systems or two resource allocation mechanisms 
for the same system one typically normalizes the social welfare metric by comparing the value to an optimal social welfare value. 
The optimal social welfare value is the value obtained if all users (in the case of no contention) or the highest paying user receive all the resources that they desire.   
Economic efficiency is defined as the optimal social welfare over the social welfare obtained using an actual allocation strategy.  
A system with an economic efficiency of 90\%, for instance have some opportunity, to allocate resource to higher paying users and 
thereby extract a higher profit. 

In essence, ensuring economic efficiency involves optimizing social welfare.

There is however an argument to be made that always allocating to the highest paying user does not 
create a healthy sustainable ecosystem, which we will discuss next.

\subsection*{Fairness}
Consider the case where some user constantly outbids a user by \$.0001 every hour in a competitive auction for resources. 
An economically efficient strategy would be to continuously allocate the resource to the highest bidder.  The bidder who keeps getting outbid 
will however at some point give up and stop bidding. This brings demand down and the resource provider may lose out on long term revenue. 
It is hence also common practice to consider the fairness of a system. In economics a fair system is a defined in terms of envy between users 
competing for the same resource~\cite{varian1974}. Envy is defined as the difference in utility that a user received for the actual allocation obtained compared 
to the maximum utility that could have been obtained across all allocations for the same resource to other users. The metric is referred to as envy-freeness and a 
fair system tries to maximize envy freeness (minimize envy). Having high fairness is important to maintain loyal customer, and it may in some cases be traded off 
against efficiency as seen in the example above. Fairness may not be efficient to obtain in every single allocation instance, 
but is commonly evaluated over a long period of time. For example a system could keep track of the fairness deficit of each user and try to 
balance it over time to allocate resources to a user that has the highest fairness deficit when resources become available. 

In addition to fairness considerations, there could be other reasons why a resource seller may want to 
diverge from a pure efficiency-optimizing strategy. If information is imperfect and the seller needs to
price goods based on the expected willingness to pay by consumers, it may be a better long-term strategy
to set the price slightly lower to avoid the dire effects of losing trades by setting the price to high. 
Another reason may be that some consumers have less purchasing power than others, and giving them
benefits, so they can stay in the market, improves the overall competitiveness (and liquidity, see below)
of the market, which in turn forces the richer consumers to bid higher.

\subsection*{Liquidity}
The central assumption in variable pricing models (see the section above on variable pricing) is that the price is a proxy or a signal for demand. 
If this signal is very accurate, allocations can be efficient and incentives to use versus back off of resources are well aligned. 
If there are too few users competing for resources the prices may plummet and the few users left may get the resource virtually for free. 
It is therefore critical for a provider to have enough competing users and to have enough purchases of resources for all the market 
assumption to come into play. In particular this means ensuring that the second part of incentive compatibility is met, i.e. 
users have an incentive to participate. Most providers fall back on fixed pricing if there is too little competition, but that may lead to 
all the inefficiency that variable pricing is designed to address. In economics this volume of usage and competition on a market is 
referred to as liquidity~\cite{keynes1965}. Lack of liquidity is a very common reason for market failure, which is why many financial and 
economic markets have automated traders to ensure that there is a trade as long as there is a single bidder who sets a reasonable price. 
A provider may, for instance, put in a daemon bidder to ensure that resources are always sold at a profit.

\subsection*{Predictability}
The biggest downside of variable pricing models is unpredictability. If the price spikes at some time in the future, 
the allocation may have to drop even though the demand is the same to avoid breaking the budget.  Exactly how much budget to 
allocate to resources depends on the predictability of the prices, i.e. the demand. If the demand is flat over time, very little excess 
budget has to be put aside to cope with situations where resources are critically needed and demand and prices are high. 
On the other hand, if some application is not elastic enough to handle resource variation, e.g. nodes being de-allocated because the 
price is too high, a higher budget may need to be allocated to make sure the application runs at some minimal level of allocation. 

Essentially users as well as applications have different sensitivity to risk of losing resource allocations or resources being more expensive.  
In economics the attitude towards risk is described in the risk-averseness or risk attitude property of a user. There are three types of users 
that differ in how much they are willing to spend to get rid of risk (variation)~\cite{pratt1964}. Risk-averse users will spend more 
money than the expected uncertain price (i.e. hedge for future spikes  c.f. the discussion on over-provisioning and under- provisioning)~\cite{arrow1965}. 
Risk-neutral users will spend exactly the expected price. Finally, risk-seekers will put in a lower budget than the expected 
price to meet their allocation needs (see Figure~\ref{fig:risk}). An application that is perfectly elastic and that may scale down or up over time as long as the long term 
performance is guaranteed may choose a risk neutral strategy. Risk seekers are less common in computational markets, but they may be 
bettering on demand going down in the future.  Risk-averse users are the most common group, and the premium they pay above 
the expected price is a good indicator for how much a resource provider can charge for reservations, which essentially 
eliminates this uncertainty.

In summary, the elasticity of a Cloud application is highly related to the risk-aversion of the resource purchase, i.e. how much to pay to hedge uncertainty.
 
\section*{Summary}
We have discussed some computational principles underlying the efficient design of Cloud computing infrastructure provisioning. 
We have also seen how economic principles play a big role in guiding the design of sustainable, profitable, and scalable systems. 
As Cloud computing becomes more commonplace and more providers enter the market, the economic principles are likely to play a bigger role. 
The sophistication of the market designs depends very much on the level of competition and usage, 
a.k.a. as the liquidity of a market.

The key to a successful market design is to align the incentives of the buyers and sellers with those of the system as a whole. 
This will ensure participation and liquidity. 
Most computational principles in the Cloud are governed by the notion that large scale distributed systems see failures so frequently that 
failover and recoverability must be an integral part of the software design.  In order to failover successfully one needs to have full 
programmatic control from hardware to end-user application. An ongoing trend has been to develop platforms and cloud operating systems 
that offer this level of software control of hardware to automate administration, management, and deployment dynamically based on demand.


\begin{backmatter}

\section*{Competing interests}
  The authors declare that they have no competing interests.

\section*{Author's contributions}
    This text is based on a distributed systems class co-taught and co-developed
    by the authors at KAIST during the fall of 2013. TS contributed mostly to the
    economics section, and DL contributed mostly to the systems section.

\section*{Acknowledgements}
   We thank the students in the class, which this text is based on, for their feedback.
   We also thank Bernardo Huberman for influencing many of the ideas presented in
   the sections on economic principles. Finally, we would also like to thank Filippo
   Balestrieri for reviewing early drafs of this letter.

\bibliographystyle{bmc-mathphys} 
\bibliography{cloud}      



\newpage
\section*{Figures}

\begin{figure}[h!]
\includegraphics[scale=0.4]{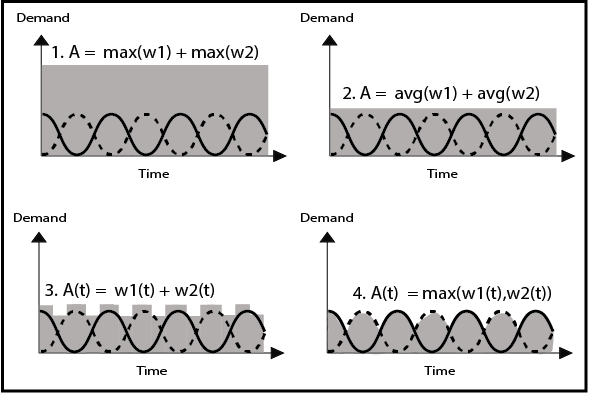}
\caption{\csentence{Statistical multiplexing.}
      Allocations for workload 1 (w1) and workload 2 (w2) competing for the same resources.}\label{fig:statmult}
\end{figure}

\begin{figure}[h!]
\includegraphics[scale=0.8]{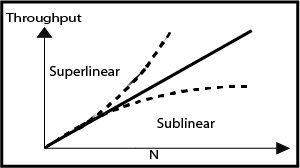}
\caption{\csentence{Scalability.}
     Superlinear - convex growth. Sublinear - concave (saturated) growth.}\label{fig:scalability}
\end{figure}

\begin{figure}[h!]
\includegraphics[scale=0.4]{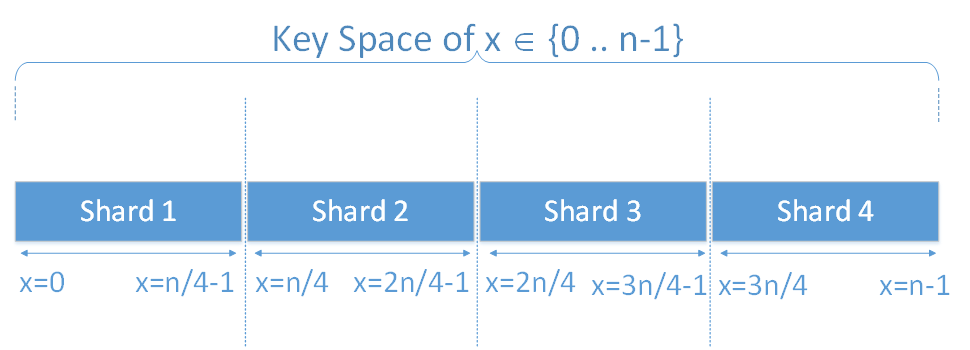}
\caption{\csentence{Range partitioning.}
     Efficient for partial key scans.}\label{fig:range}
\end{figure}

\begin{figure}[h!]
\includegraphics[scale=0.4]{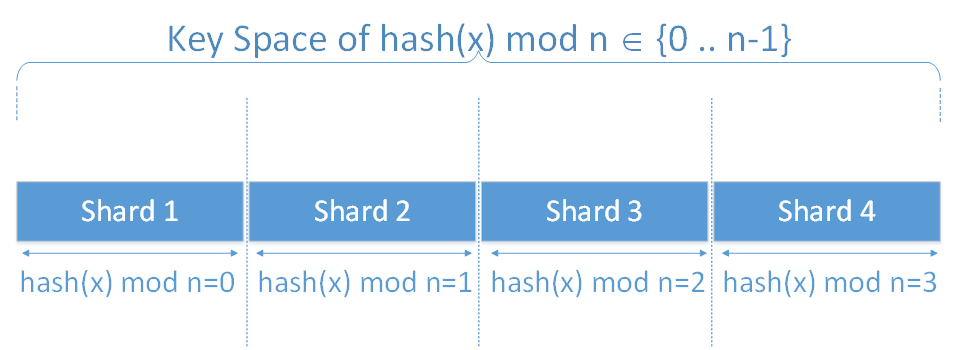}
\caption{\csentence{Hash partitioning.}
     Efficient for partial key scans.}\label{fig:hash}
\end{figure}

\begin{figure}[h!]
\includegraphics[scale=0.4]{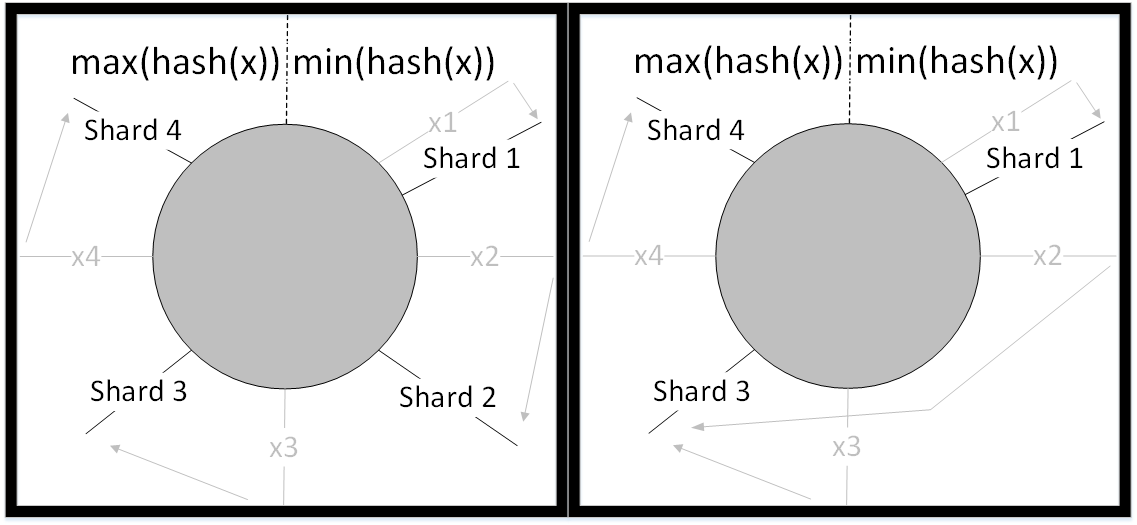}
\caption{\csentence{Consistent hashing.}
     x1..x4 denote keys. If the Shard 2 machine goes down only key x2 needs to be reassigned.}\label{fig:consist}
\end{figure}

\begin{figure}[h!]
\includegraphics[scale=0.8]{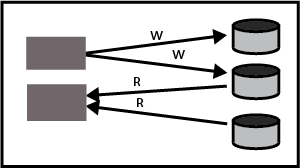}
\caption{\csentence{Eventual consistency.}
     Example fault-tolerant configuration.}\label{fig:eventual}
\end{figure}

\begin{figure}[h!]
\includegraphics[scale=0.8]{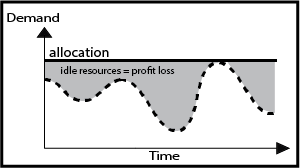}
\caption{\csentence{Over provisioning.}
     Profit opportunities are lost due to many idle resources.}\label{fig:over}
\end{figure}

\begin{figure}[h!]
\includegraphics[scale=0.8]{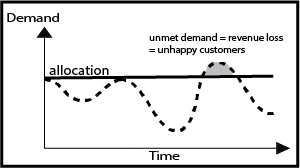}
\caption{\csentence{Under provisioning.}
     Demand is unmet and therefore revenue opportunities are lost.
     Service downtime may also lead to long-term revenue loss due to lost customers.}\label{fig:under}
\end{figure}

\begin{figure}[h!]
\includegraphics[scale=0.8]{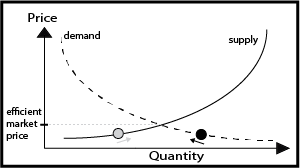}
\caption{\csentence{Supply and demand curves.}
     The efficient market price is where the supply and demand curves meet.
     Pricing below may lead to shortage of supply. Increasing the price towards the market price
     will take the demand down to a point were it can be met.}\label{fig:supply}
\end{figure}

\begin{figure}[h!]
\includegraphics[scale=0.8]{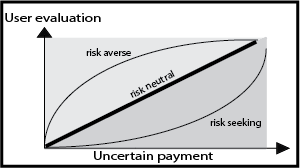}
\caption{\csentence{Risk attitudes.}
     Risk averseness is the amount of money you are willing to pay to remove risk.
     Risk neutral people will always be willing to pay the same amount for a lottery ticket
     as the expected outcome or gain of the lottery.} \label{fig:risk}
\end{figure}

\end{backmatter}
\end{document}